# Funding information in Web of Science: An updated overview

**Forthcoming in *Scientometrics***

Weishu Liu
wsliu08@163.com
School of Information Management and Engineering, Zhejiang University of Finance and Economics, Hangzhou 310018, Zhejiang, China

Li Tang
litang@fudan.edu.cn
School of International Relations and Public Affairs, Fudan University, Shanghai 200433, China

Guangyuan Hu* Corresponding author
hu.guangyuan@shufe.edu.cn
School of Public Economics and Administration, Shanghai University of Finance and Economics, Shanghai 200433, China

**Abstract:** Despite the limitations of funding acknowledgment (FA) data in Web of Science (WoS), studies using FA information have increased rapidly over the last several years. Considering this WoS' recent practice of updating funding data, this paper further investigates the characteristics and distribution of FA data in four WoS journal citation indexes. The research reveals that FA information coverage variances persist cross all four citation indexes by time coverage, language and document type. Our evidence suggests an improvement in FA information collection in humanity and social science research. Departing from previous studies, we argue that FA text (FT) alone no longer seems an appropriate field to retrieve and analyze funding information, since a substantial number of documents only report funding agency or grant number information in respective fields. Articles written in Chinese have a higher FA presence rate than other non-English WoS publications. This updated study concludes with a discussion of new findings and practical guidance for the future retrieval and analysis of funded research.

**Acknowledgment:**
This research draws on support from the National Natural Science Foundation of China (#71801189), the Ministry of Education of China (#18YJAZH027), and the Natural Science Foundation of Zhejiang Province (#LQ18G030010). We would like to express our deep thanks to two anonymous reviewers for their comments and suggestions. The conclusions contained herein are those of the authors and do not reflect the views of the funders.

## 1. Introduction

Public resources are scarce, as are research grants. With more economies escalating their investments in basic research, the effectiveness and efficiency of public funding has attracted considerable attention from both researchers and policymakers (Liu et al., 2019; Jacob & Lefgren, 2011; Ubfal & Maffioli, 2011; Yin et al., 2018). Yet, due to the absence of large-scale micro-level data, as noted by Moller et al. (2016), important questions such as the effect of research funding on scientific output and the strategic funding portfolios of different entities, are difficult to investigate empirically, with very few exceptions. For example, based on a selection of UK papers indexed in the Social Sciences Citation Index (SSCI) and Science Citation Index-Expanded (SCIE), Lewison and Carding (2003) discovered that approximately 50% of British speech and language research had no funding acknowledgement (FA).

In August 2008, Web of Science (WoS) began to systematically collect FA data.[1] This valuable data introduces a new venue of various funding-related studies. For instance, using FA data in nanotechnology publications, Shapira and Wang investigated the effects of funding on the development trajectory of then-emerging research (Shapira & Wang, 2010; Wang & Shapira, 2011; Wang & Shapira, 2015). According to a search in WoS, 193 indexed articles explored this topic from 2016 to 2018.[2] In the field of information and library science, prior studies have investigated funding patterns in different scientific domains (Álvarez-Bornstein et al., 2019; Liu et al., 2015; Mejia & Kajikawa, 2018; Paul-Hus et al., 2017a; Zhao et al., 2016) and across various economies (Alvarez & Caregnato, 2018; Huang & Huang, 2018; Wang et al., 2012). Some studies rely solely on funding data to explore the characteristics of funded research or the effects of funding (Gök et al., 2016; Costas & van Leeuwen, 2012; Yan et al., 2018; Zhao et al., 2018). Others combine FA with other complementary information such as authorship or authors' notes to investigate the structure of the division of labor in the scientific community (Walsh et al., 2019), the partnership between the public and private sectors (Morillo, 2016), and explicit and implicit collaboration patterns (Paul-Hus et al., 2017b).

Data quality is the cornerstone of empirical studies. Along with the increasing numbers of studies based on FA data, some researchers have also investigated the limitations of data provided by WoS (Franceschini et al., 2016; Liu, 2019; Liu et al., 2018b; Zhu et al., 2019a, 2019b), including funding data (Lundberg et al., 2006; Paul-Hus et al., 2016; Tang et al., 2017). Some aspects of funding data such as completeness, accuracy, and funding agency variation have been well documented (Alvarez-Bornstein et al., 2017; Grassano et al., 2017; Morillo & Alvarez-Bornstein, 2018; Powell, 2019; Tang, 2013). Others further explore the coverage biases of FA information in language, document type, and research fronts. For instance, Paul-Hus et al. (2016) note that of the three core journal citation indexes, the Arts & Humanities Citation Index (A&HCI) is not suitable for funding analysis, while the SSCI is suitable for publications indexed after 2014. In Tang et al.'s (2017) critique of FA information, they point out that, in addition to the pitfalls of name variants, name misspellings, and inaccurate FA information, there are inherent and heterogeneous biases in WoS' approach to collecting FA information. Winkelman and Rots (2016) demonstrate that a sole reliance on grant acknowledgements will lead a researcher to vastly underestimate the research outputs attributable to the observatory. These previous explorations remind bibliometricians of these potential caveats in applying FA data for funding analyses.

A review of extant studies on FA information suggests a number of research gaps. For instance,

---

[1] For more information, see http://wokinfo.com/products_tools/multidisciplinary/webofscience/fundingsearch/.
[2] The search query we use on the WoS advanced search platform is: TS= ((fund or funded or funding) AND ("Web of Science" OR WoS)), Indexes = SCIE/SSCI, Timespan = 2016–2018. Data accessed on July 8, 2019. Unless otherwise specified, all document types are considered. All data used in this paper were retrieved in June 2019 via the library of Xi'an Jiao Tong University and crosschecked at the Fudan University Library.

we are not aware of any studies that have explored FA information in the Emerging Sources Citation Index (ESCI) dataset. It is also important to point out that Clarivate Analytics is updating its FA information collection, which has not been explored in scholarly research. As stated in its online help file:

> In 2016 Web of Science Core Collection began supplementing the grant information with grant agencies and grant numbers from MEDLINE and researchfish®. Records that already contained grant information will not be changed. Records that did not have grant information were updated with grant information from MEDLINE and researchfish®.[3]

Yet it remains unclear how this new practice affects the collection of funding data in different citation indexes and retrieval practices, if at all. Also, given the increasing importance and use of funding data in bibliometrics and science policy research, we argue that it is necessary and timely to conduct an updated study on funding data in WoS.

We explore the FA information in four journal citation indexes of the WoS Core Collection: SCIE, SSCI, A&HCI, and ESCI.[4] Funding agency (FO), funding grant number (FG) and FA text (FT) are three funding-related fields provided in WoS to retrieve FA information (Rigby, 2011). Search queries #1 through #3 are applied, respectively, on the WoS advanced search platform to provide a general idea of how many records have funding information[5]. Although WoS has systematically collected FA data since August 2008, our prior experiments revealed that substantial numbers of WoS records contained FA information prior to 2008. In order to give a more dynamic and comprehensive picture of the FA data provided by WoS, in this study we extend the time span to the period of 2000–2018 for trend analysis.

**Search Query #1:** FT=(A* OR B* OR C* OR D* OR E* OR F* OR G* OR H* OR I* OR J* OR K* OR L* OR M* OR N* OR O* OR P* OR Q* OR R* OR S* OR T* OR U* OR V* OR W* OR X* OR Y* OR Z* OR 0* OR 1* OR 2* OR 3* OR 4* OR 5* OR 6* OR 7* OR 8* OR 9*)

**Search Query #2:** FO=(A* OR B* OR C* OR D* OR E* OR F* OR G* OR H* OR I* OR J* OR K* OR L* OR M* OR N* OR O* OR P* OR Q* OR R* OR S* OR T* OR U* OR V* OR W* OR X* OR Y* OR Z* OR 0* OR 1* OR 2* OR 3* OR 4* OR 5* OR 6* OR 7* OR 8* OR 9*)

**Search Query #3:** FG=(A* OR B* OR C* OR D* OR E* OR F* OR G* OR H* OR I* OR J* OR K* OR L* OR M* OR N* OR O* OR P* OR Q* OR R* OR S* OR T* OR U* OR V* OR W* OR X* OR Y* OR Z* OR 0* OR 1* OR 2* OR 3* OR 4* OR 5* OR 6* OR 7* OR 8* OR 9*)

This rest of the paper is organized as follows. After delineating the data sources and method used in this study, we profile, replicate, and compare the general pattern of FA information in WoS via different search queries. Next, we probe in detail the discrepancies in the search results from different citation index datasets over time. We then further differentiate the funding information by publication year, document type, and language. The paper concludes with a brief discussion of the new FA data findings and future research directions.

---

[3] For more detail, please refer to https://images.webofknowledge.com/images/help/WOS/hp_full_record.html.
[4] As noted in Clarivate Analytics, ESCI indexes regional journals in natural sciences, social sciences, and arts and humanities.
[5] If any of the above three funding-related fields of one record in WoS is NOT empty, the research document is deemed funded.

## 2. Analysis
### 2.1 General patterns and new updates of FA information
Table 1 lists the retrieved hits with the above three search queries and their Boolean combinations.

**Search Query #4**: (Search Query #1) OR (Search Query #2) OR (Search Query #3)
**Search Query #5**: (Search Query #2) OR (Search Query #3)
**Search Query #6**: ((Search Query #2) OR (Search Query #3)) NOT (Search Query #1)
**Search Query #7**: (Search Query #1) NOT ((Search Query #2) OR (Search Query #3))

Our analysis reveals two main findings. First, echoing previous studies (Paul-Hus et al., 2016; Tang et al., 2017), the discrepancies of returned hits when searching in different funding information fields remain. This is unsurprising, as journal articles may only report funding agency information without the funding grant number, or vice versa. It is also possible that research articles report funded project names but no grant numbers or funding agency names. Additionally, WoS' new practice of supplementing FA data with grant agencies and grant numbers retrieved from MEDLINE and Researchfish adds more information to the FO and FG fields. All of these can account for the differences in retrievals when the FT, FO, and FG fields are searched separately.

Second, and more importantly, in contrast to previous studies, we find that FT alone is no longer the ideal field tag to retrieve FA information. Both Paul-Hus et al. (2016) and Tang et al. (2017), using Center for Science and Technology Studies in-house WoS data and online FA information retrieval, respectively, conclude that the content contained in the FT field is more comprehensive than that listed in the FO and FG fields, and that the number of retrieved hits via FT captured almost all retrievals via the FO or FG fields.[6]

However, this is no longer the case, as the total number of records with FT information is smaller than that with FO information. This suggests that a large number of records (i.e., 231,000 funded articles during 2009–2018 and over 848,000 funded articles published between 2000 and 2018) will be missed when searching in the FT field tag.

To better understand this change, we also replicate and list the retrieval of Tang et al. (2017) using the same search query; the results are listed in the last two columns of Table 1.[7] This replication reveals that, contrary to previous findings, there are more FO-retrieved records than those searched from the FT field. On the one hand, this supports Clarivate Analytics' claim that they have updated their FA collection practice, while on the other hand it demonstrates the need to better understand the new characteristics of FA information over time and across different citation indexes.

---

[6] In Tang et al.'s (2017) analysis conducted in 2015, only four out of 4.6 million publications contained information in the FO or FG but not the FT field. Thus, they suggested using the FT to retrieve WoS FA information.
[7] Paul-Hus et al. (2016) analyzes FA information using the in-house Leiden data. Due to data accessibility issues, we could not replicate and compare their retrieved hits with various combinations of searches in our paper.

Table 1 Search results of funding information by field, citation index and retrieval time

| Search Query# | Search fields | Num. Rec (New) | Num. Rec (New) | Tang et al. 2017 | |
|---|---|---|---|---|---|
| | | | | Num. Rec (Original) | Num. Rec (Replication) |
| 1 | FT | 9,694,586 | 9,464,482 | 4,610,481 | 4,630,078 |
| 2 | FO | 10,499,161 | 9,653,150 | 4,591,259 | 4,769,283 |
| 3 | FG | 7,859,126 | 7,111,055 | 3,171,084 | 3,380,809 |
| 4 | FT OR FO OR FG | 10,542,655 | 9,695,711 | 4,610,485 | 4,788,930 |
| 5 | FO OR FG | 10,542,552 | 9,695,614 | 4,610,387 | 4,788,842 |
| 6 | (FO OR FG) NOT FT | 848,069 | 231,229 | 4 | 158,852 |
| 7 | FT NOT (FO OR FG) | 103 | 97 | 98 | 88 |
| **Time period** | | Y00-Y18 | Y09-Y18 | Y09-Y14 | Y09-Y14 |
| **Examined datasets** | | SCIE, SSCI, A&HCI, ESCI | SCIE, SSCI, A&HCI, ESCI | SCIE, SSCI, A&HCI | SCIE, SSCI, A&HCI |
| **Retrieval time** | | June 2019 | October 2019 | December 2015 | June 2019 |

*Note*: All document types considered.

*2.2 Time dynamics of the funding information by dataset*

We study the years 2000–2018 to depict the dynamics for SCIE, SSCI, and A&HCI. As the ESCI dataset is available from 2015, its entire duration (2015–2018) is chosen for further analysis. We start with search query #4 to provide a general idea of the number of indexed publications reporting funding information in each citation index dataset. Figure 1 illustrates the temporal analysis results. The grey lines indicate the funded records in each dataset, and the bars represent the percentage of records reporting funding information. To further understand which funding-related search fields contribute to the newly added information, we divide funded publications into two mutually exclusive subsets: records with FT information (FT group, blue bars) and records without FT information but with information on funding agency or funding grant number ((FO+FG-FT) group, orange bars).[8]

*SCIE dataset*

Panel A shows the time dynamics for SCIE. Unsurprisingly, 2008 is the watershed year for the presence of FA information. About one-fifth of records published in 2008 report funding information. This rate jumped to 40% in 2009 and gradually increased to 58% in 2018. Two new features of FA information are revealed in Panel A. First, in contrast to Paul-Hus et al.'s (2016) finding that almost no FA information is available for WoS publications prior to 2008, we find that 5% of all publications indexed in the SCIE reported funding information from 2000 to 2007. Second, in contrast to Tang et al.'s results (2017), our data shows that WoS' practice of updating FA information renders the FT field alone no longer ideal for retrieving FA information. During the period of 2009–2018, approximately 175,000 SCIE records belonged to the (FO+FG-FT) group. From 2000 to 2007, the (FO+FG-FT) group takes the dominant share: 521,000 records, i.e. nearly 96% of all SCIE-indexed funded publications, do not report funding information in the FT field. These results suggest that only using the FT field for FA analysis will miss a substantial number of funded records.

---

[8] The blue bars in Figure 1 represent the returned hits of search query 1, denoted [#1]; orange bars represent the returned hits of search query 6 ([#6]). As shown in Table 1, the returned hits of search query 4 ([#4]) equal the sum of [#1] and [#6] or the sum of [#5] and [#7]. So, another division can be recorded with information in FO or FG ([#5]), and records with information in FT but without data in FO or FG ([#7]). However, as [#7] can have as few as 103 results, we only use the subsets of [#1]+[#6].

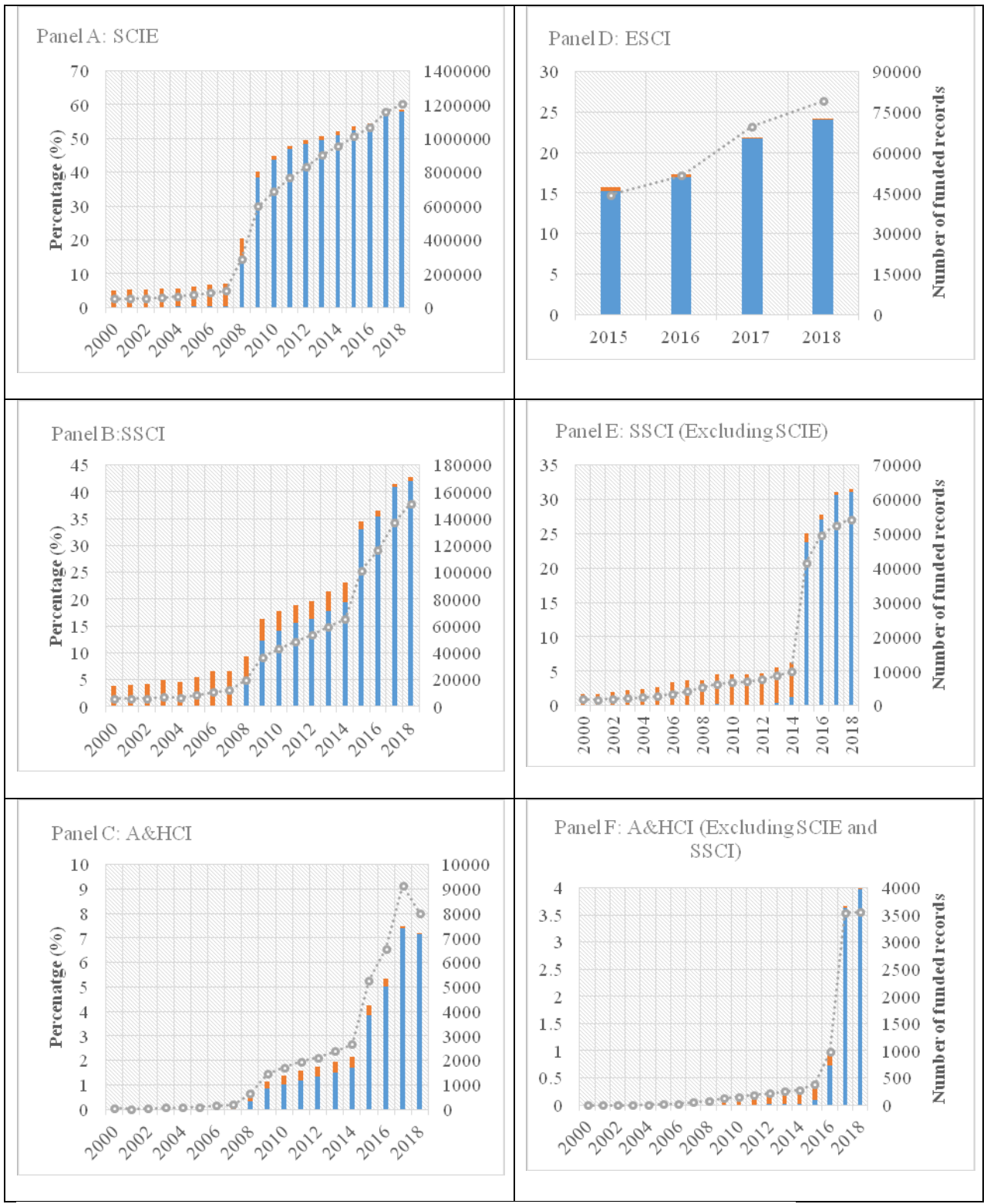

Figure 1 Time dynamics of funding reporting rates in citation index datasets

*SSCI dataset*
In a similar vein, FA information provided in the SSCI (Figure 1, Panel B) displays a pattern that is different than that observed in previous studies (Paul-Hus et al. 2016; Tang et al. 2017). Approximately 63,000 SSCI publications (5% of the records) published from 2000 to 2007 contained funding information. The funding reporting rates increased gradually after 2008. From 2000 to 2007, nearly 99% of funded records belong to the (FO+FG-FT) group; very few records can be found via searching the FT field. The pattern begins changing in 2008, when the (FO+FG-FT) and FT groups report 56% and 44% the funded publications, respectively.

*A&HCI dataset*
We observe a similar pattern of funding information distribution in the A&HCI dataset. For funded A&HCI records published in 2000–2007, a majority were only reported in (FO+FG-FT) groups. Starting in 2009, an increasing share of FA information was also available in the FT field, but still approximately 3,885 funded research projects are only available in the FG or FO fields during 2009–2018. Compared to the SCIE and SSCI datasets, the presence of funding rates in the A&HCI records is much lower. As shown in Figure 1, Panel C, between 2000 and 2007 fewer than 0.2% of the A&HCI records report funding information. The funding reporting rate begins to rise slowly from 0.5% in 2008 to 4.2% in 2015 and 7.5% in 2017.

*ESCI dataset*
In addition to the three flagship citation indexes (SCIE, SSCI, A&HCI), we also examine the pattern of FA information in the ESCI, a new supplementary journal citation index that focuses on journals of regional importance and in emerging fields (Huang et al., 2017). We studied the period 2015–2018 because the dataset was launched in 2015.[9] Panel D of Figure 1 illustrates that the funding reporting rates in ESCI records range between 15% and 24% with an upward trend. Similar to the SCIE, SSCI, and A&HCI, the FT field captures a large share of funded ESCI records during the years 2015–2018. However, 3,566 records (about 1.5% of all funded records) report only the funding agency or funding grant number but not FT information. If we rely only on the FT field, these records will be also missed.

*SSCI (excluding the SCIE) and A&HCI (excluding the SCIE and SSCI)*
Tang et al. (2017) found that FA information collected in WoS core datasets between 2009 and 2014 was mainly contributed by journals indexed in the SCIE. Paul-Hus et al. (2016) reported that no FA information was collected for records published in 2015 and solely indexed in the A&HCI. To further examine if these patterns still hold with WoS' updating FA collection, Panels E and F in Figure 1 plot the dynamics of FA presence for the SSCI (excluding the SCIE) and A&HCI (excluding the SCIE and SSCI), respectively.

The key message comparing Panels B and E, and Panels C and F, is that there has been a significant improvement in FA information collection in the SSCI and A&HCI datasets. While journals indexed in the SCIE remain the main contributor of FA information in SSCI and A&HCI records, an increasing number of articles only indexed in the SSCI or A&HCI have FA information available; in 2018 these figures reached 54,000 (Panel E) and 3,500 (Panel F), respectively. Second, consistent with the findings presented in Panels A through D, the FT field alone is not the best way to retrieve FA information. For the SSCI (excluding the SCIE), 95% of these funded records only have funding agency or funding grant number information recorded from 2000 to 2014. For the A&HCI (excluding the SCIE and SSCI), 90% of these funded records only have funding agency or funding grant number information recorded from 2000 to 2015.

---

[9] For more detailed information, see https://clarivate.com/essays/journal-selection-process/

In summary, the first message conveyed by Figure 1 is that, in all citation indexes examined, both the absolute quantity and relative share of articles reporting funding have grown over time, with few exceptions.[10] Second, the first full year in which FA information was reported varies in WoS' different citation indexes. FA data have been systematically collected for SCIE-indexed articles since 2009, SSCI from 2015, A&HCI from 2017, and the ESCI from the first year it was available (2015).[11] Third, among the three core citation indexes (SCIE, SSCI and A&HCI), the FT field alone is not suitable for retrieving funding information. This is true for the whole period of 2009–2018, but particularly prior to 2008. One salient pattern illustrated in six panels of Figure 1 is that many records consistently contain information on funding agency or funding grant number but not FT information (i.e. the retrievals of search query #6), which suggests that searching the FT field alone is not effective for comprehensive FA information retrieval. And finally, compared to previous studies, our data reveals that FA information collection has made substantial improvements in social science and humanities research in recent years.

*2.3 Document type and the presence of funding information*
In the above two sections, all the document types in WoS are considered. Yet in bibliometric analysis, some document types such as news items weigh less in terms of originality than others. Thus, in this section we focus on four more heavily weighted document types—article, editorial, letter, and review—and further examine their respective funding reporting rate. Since only 103 records reported FT but not funding agency or grant numbers from 2000 to 2018, we use the FO or FG fields to analyze the presence of FA information.

Figure 2 shows the funding reporting rates by the four document types over time. We set the starting year for each dataset based on when it began to systematically collect funding information: 2009 for SCIE, 2015 for SSCI, 2017 for A&HCI, and 2015 for ESCI.

The analysis reveals three main findings. First, not only were the funding text of articles and reviews published before 2016 collected (Paul-Hus et al. 2016), this funding information is now also reported and collected for editorials and letters. Second, funding reporting rates are much higher for original articles and reviews than for editorials and letters. This finding is consistent across all four citation index datasets. For example, funding information was included in only 2.5% of editorials listed in the A&HCI dataset for 2017–2018; likewise, only 2 out of 2,859 letters reported funding information in 2017–2018. This might due to the length and nature of editorials and letters, there is no requirement for authors to submit funding information. Third, the presence of FA information is lowest in the A&HCI in both absolute numbers and relative share for all four types of publications. This could be due to inconsistent indexing/retrieval practices of WoS data across different citation indexes. Or it may reflect the common practice of humanity journals not requiring FA information reporting. It is also plausible that it is simply an artifact of comparatively less financial support being provided to humanities research.

---

[10] For instance, the presence of FA information in the A&HCI dataset is higher for 2017 than for 2018.
[11] It should be noted that about 600 journals are both SSCI and SCIE indexed according to the 2018 Journal Citation Reports, suggesting that financial support information of SSCI publications in 2010 can be identified and retrieved if they are also indexed in the SCIE.

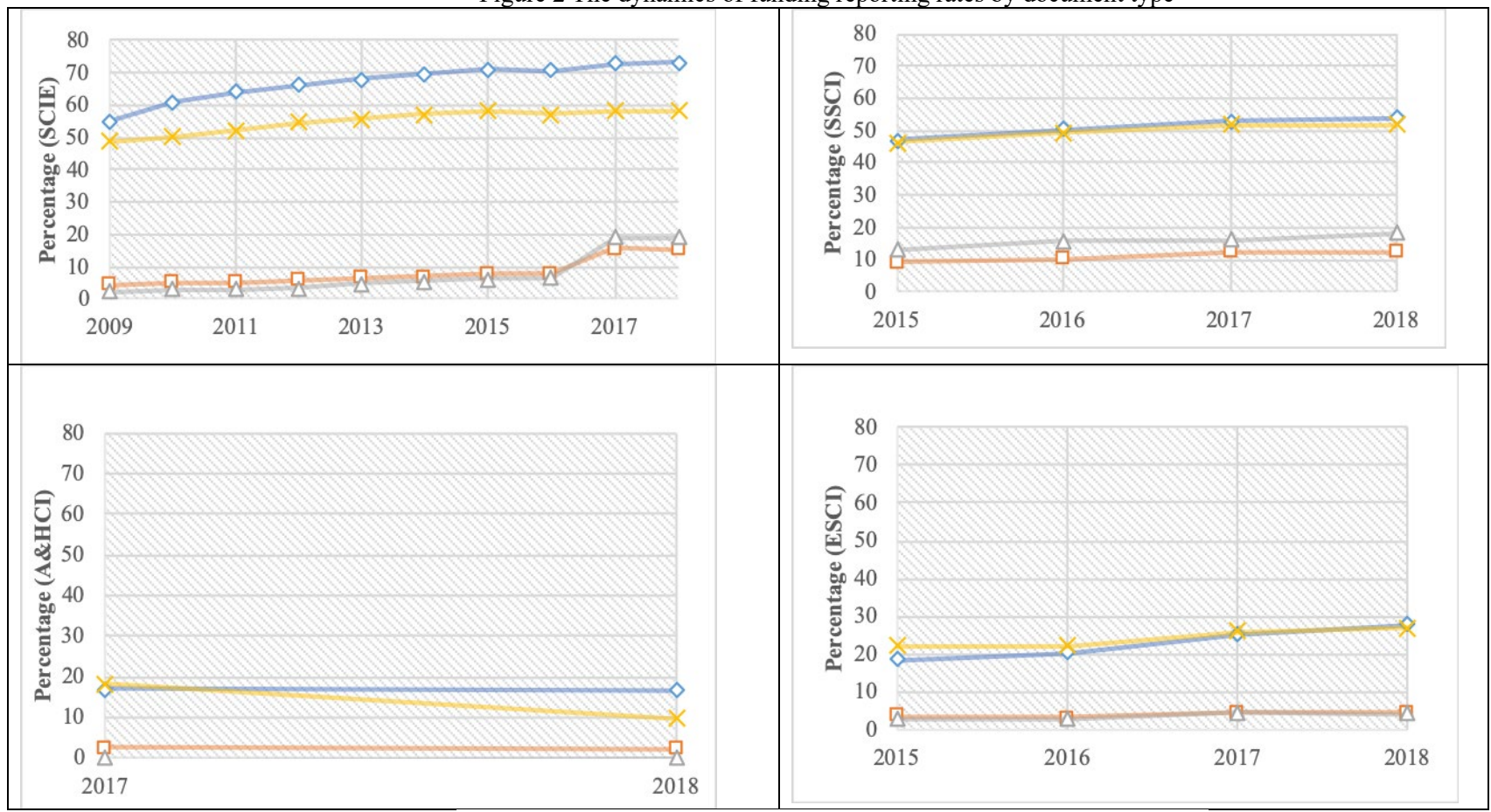

Figure 2 The dynamics of funding reporting rates by document type

*2.4 FA information presence by language*

English is the lingua franca of global scholarly communication. It is well known that WoS indexes a disproportionately large share of records published in English (Liang, Rousseau, & Zhong, 2013; Liu, 2017; Liu et al., 2018a; Mongeon & Paul-Hus, 2016). Non-English publications are only eligible for WoS indexation when their title, abstract and bibliographic information are available in English. This also applies to FA information.[12] Tang et al. (2017) pointed out that SCIE records prior to 2015 have much higher funding reporting rates for articles published in English and Chinese. Given the changing presence of FA information in the SSCI and A&HCI databases, we expand Tang et al.'s study, which only analyzed the SCIE, to the other three citation indexes. We are curious whether the new updates of funding information reveal any changes in the distribution of publishing languages in the four core citation indexes of WoS. The study periods for the four indexes are set as described above.

*SCIE dataset*

The data shows that about 17.8 million records were published in the SCIE from 2009 to 2018, of which 97.5% of the articles were written in English, the universal language of science publishing. During this period, 9.18 million SCIE records were funded, of which 99.7% were published in English. Table 2 lists the top 10 languages of SCIE records published from 2009 to 2018. The table contains the same languages as for 2009–2014 (Tang et al., 2017), except French and Spanish swapped third and fourth places. Consistent with the findings reported by Tang et al. (2017), English and Chinese have the highest funding reporting rates. Other main publishing languages such as German, Japanese, French, and Russian have extremely low funding reporting rates.

Table 2 Funding reporting rates of the main publishing languages in the SCIE (2009–2018)

N=17,840,472

| Language | #Rec in SCIE | Percent of SCIE | #Rec with FA | FA presence rate % | Language share of funded SCIE |
|---|---|---|---|---|---|
| ***English*** | *17,392,319* | *97.49* | *9,158,220* | *52.66* | *99.70* |
| German | 127,225 | 0.71 | 141 | 0.11 | 0.00 |
| Spanish | 69,855 | 0.39 | 285 | 0.41 | 0.00 |
| French | 67,466 | 0.38 | 157 | 0.23 | 0.00 |
| ***Chinese*** | *62,213* | *0.35* | *25,708* | *41.32* | *0.28* |
| Portuguese | 41,954 | 0.24 | 116 | 0.28 | 0.00 |
| Polish | 17,168 | 0.10 | 28 | 0.16 | 0.00 |
| Japanese | 13,765 | 0.08 | 21 | 0.15 | 0.00 |
| Russian | 10,813 | 0.06 | 21 | 0.19 | 0.00 |
| Turkish | 7,719 | 0.04 | 44 | 0.57 | 0.00 |
| ***Top 10 subtotal*** | ***17,810,497*** | ***99.83*** | ***9,184,741*** | ***51.57*** | ***99.98*** |

*SSCI dataset*

Between 2015 and 2018, approximately 1.29 million publications were indexed in the SSCI. Of these, 96.9% were published in English, followed by Spanish (0.93%) and German (0.93%). On average, 40% of SSCI records written in English reported their funding information, while more than 99.9% of the funded SSCI records were published in English.

---

[12] For details about WoS policy on journal indexing, see
https://support.clarivate.com/ScientificandAcademicResearch/s/article/Web-of-Science-Core-Collection-Submission-and-indexing-of-Journals-not-written-in-English-Language?language=en_US.

Table 3 lists the top 10 most frequent publishing languages in SSCI publications, the number of funded records published in those languages, and funding reporting rates.

The table shows that the funding reporting rate for Russian documents is 8.4%, much higher than those of the other eight main publishing languages. Chinese-language documents, in sharp contrast to their position in SCIE publications, are squeezed out of the top 10 most frequent languages for SSCI papers. Over the period 2015–2018, Chinese is ranked the 19[th] most frequent language in the SSCI with 78 articles, in which 25 report FA information with a high FA presence rate of 32%. This echoes the finding by Flowerdewa and Li (2009) that, in humanities and social sciences, Chinese scholars' international publications continue to increase in English. However, this supports the appeals by some Chinese scholars to construct academic discourses of power if China aims to become an Olympic player and exert more influence in the realm of social studies (Nie, 2014; Wu et al. 2011; Xie, 2013;).

Table 3 Funding reporting rates of the main publishing languages in the SSCI (2015–2018)

N=1,297,058

| Language | #Rec in SSCI | Percent of SSCI | #Rec with FA | FA presence rate % | Language share of funded SSCI |
|---|---|---|---|---|---|
| ***English*** | *1,257,038* | *96.91* | *506,021* | *40.26* | *99.92* |
| Spanish | 12,068 | 0.93 | 67 | 0.56 | 0.00 |
| German | 12,037 | 0.93 | 38 | 0.32 | 0.00 |
| French | 5,199 | 0.40 | 8 | 0.15 | 0.00 |
| Portuguese | 3,463 | 0.27 | 8 | 0.23 | 0.00 |
| ***Russian*** | *1,840* | *0.14* | *154* | *8.37* | *0.00* |
| Italian | 807 | 0.06 | 2 | 0.25 | 0.00 |
| Czech | 780 | 0.06 | 1 | 0.13 | 0.00 |
| Dutch | 640 | 0.05 | 1 | 0.16 | 0.00 |
| Turkish | 562 | 0.04 | 4 | 0.71 | 0.00 |
| ***Top 10 subtotal*** | *1,294,434* | *99.79* | *506,304* | *39.11* | *99.97* |

*A&HCI dataset*
The publishing languages of A&HCI indexed records are more diverse than those in the SCIE and SSCI indexes. About 76.9% of A&HCI records were published in English in 2017 and 2018, followed by French (7.4%), German (5.4%), and Spanish (3.9%).

As depicted in Table 4, less than one-tenth of English-language A&HCI records published in 2017 and 2018 report funding information, followed by Russian documents, with a 2.2% funding report rate. However, the publishing languages of funded A&HCI records are still concentrated: more than 99% of the funded A&HCI records are published in English. Over the last two years, only seven out of 544 Chinese-language A&HCI articles reported funding information.

*ESCI index*
English is again unquestionably the dominant written language in the ESCI: 76.7% of the records published from 2015 to 2018 are in English. Like the A&HCI dataset, the publishing languages of ESCI records are also dispersed. After English, Spanish (7.9%), Russian (3.6%), and Portuguese (3.1%) are the main publishing languages of ESCI records. However, English (97%) is the dominant publishing language for funded ESCI records published from 2015 to 2018, followed by Chinese (1.7%) and Russian (1.2%).

Table 4 Funding reporting rates of the main publishing languages in the A&HCI (2017–2018)
N=233,554

| Language | #Rec in A&HCI | Percent of A&HCI | #Rec with FA | FA presence rate % | Language share of funded A&HCI |
|---|---|---|---|---|---|
| English | 179,630 | 76.91 | 17,038 | 9.49 | 99.28 |
| French | 17,347 | 7.43 | 3 | 0.02 | 0.02 |
| German | 12,559 | 5.38 | 6 | 0.05 | 0.03 |
| Spanish | 9,080 | 3.89 | 18 | 0.20 | 0.10 |
| Italian | 6,117 | 2.62 | 5 | 0.08 | 0.03 |
| Russian | 2,757 | 1.18 | 60 | 2.18 | 0.35 |
| Portuguese | 1,186 | 0.51 | 1 | 0.08 | 0.01 |
| Czech | 786 | 0.34 | 1 | 0.13 | 0.01 |
| Dutch | 765 | 0.33 | 0 | 0.00 | 0.00 |
| Chinese | 544 | 0.23 | 7 | 1.29 | 0.04 |
| *Top 10 subtotal* | *230,771* | *98.81* | *17,139* | *7.43* | *99.87* |

Table 5 lists the top 10 publishing languages and their corresponding numbers of funded records and funding reporting rates for the ESCI index. For English ESCI records, 25% report funding information. Surprisingly, 61% of the ESCI records published in Chinese report funding information, which is much higher than the funding reporting rates for Chinese-language publications in the SCIE, SSCI and A&HCI datasets. ESCI records published in Korean and Russian also demonstrate a moderate rate of funding reporting.

Table 5 Funding reporting rates of the main publishing languages in the ESCI (2015–2018)
N=1,225,685

| Language | #Rec in ESCI | Percent of ESCI | #Rec with FA | FA presence rate % | Language share of funded ESCI |
|---|---|---|---|---|---|
| *English* | *940,183* | *76.71* | *236,367* | *25.14* | *96.59* |
| Spanish | 96,837 | 7.90 | 69 | 0.07 | 0.03 |
| Russian | 43,846 | 3.58 | 2,856 | 6.51 | 1.17 |
| Portuguese | 38,477 | 3.14 | 44 | 0.11 | 0.02 |
| German | 23,271 | 1.90 | 128 | 0.55 | 0.05 |
| French | 20,763 | 1.69 | 47 | 0.23 | 0.02 |
| Italian | 12,039 | 0.98 | 54 | 0.45 | 0.02 |
| Turkish | 10,413 | 0.85 | 47 | 0.45 | 0.02 |
| Korean | 6,696 | 0.55 | 859 | 12.83 | 0.35 |
| *Chinese* | *6,623* | *0.54* | *4,042* | *61.03* | *1.65* |
| *Top 10 subtotal* | *1199148* | *97.83* | *244513* | *20.39* | *99.92* |

## 3. Conclusion and discussion

Donors, funders, and science policy makers around the world are increasingly championing evidence-based research evaluation. By examining the characteristics and distribution of the most recent FA information in four journal citation indexes of WoS, this study provides a timely update to previous studies on WoS FA information (Paul-Hus et al., 2016; Tang et al., 2017). We find that in all four citation indexes, both the absolute number and relative share of articles reporting funding information have increased over the years. This improvement is particularly obvious for earlier years and for social science and humanities research.

English-language publications are consistently the most comprehensively covered in FA information records across all journal citation indexes; WoS records written in other languages are far less likely to contain FA details. Articles written in Chinese are much more likely to contain FA information than other non-English WoS publications. We also find that a substantial number of funded records in WoS have funding information available in FO and FG only, which suggests that the previously recommended data retrieval approach is not suitable for online searches for FA data.

Unquestionably, WoS' retroactive efforts to add FA information for previously indexed publications indicate that this information has a large untapped potential in both research assessment and science policy evaluation. In future research, analyses that characterize funded research by institution, country, discipline as well as funding source type (such as international co-funding or industrial funding) would be interesting to explore over a longer time period. The starting point of any rigorous research assessment and evidence-based policy evaluation is credible and high-quality data. We therefore propose the following three-step procedure for future research using FA information in WoS.

Step 1 is to decide whether WoS FA information is the appropriate data to address the research question. In addition to the well-known coverage biases of datasets (Liang et al., 2013; Liu, 2017; Mongeon & Paul-Hus, 2016), the results show that these biases in FA information persist in all four citation indexes to varying extents by time coverage, language and document types. For example, the FA data in WoS have been more systematically collected for the SCIE since 2009, for the SSCI since 2015, for the A&HCI since 2017, and for the ESCI since 2015. This suggests that studies aiming to profile funding patterns or evaluate funding impacts written in languages other than English should not rely on FA information in WoS.

Step 2 is how to retrieve credible data. The most comprehensive approach to retrieving FA information is to use search query #4 to combine information on all funding-related fields (FO, FG and FT). Given that only a very limited share of funded records will be missed by FO and FG, we argue that search query #5 should be employed to increase efficiency. Finally, step 3 should discuss the caveats of the potential influence of the quality of the FA information provided to remind readers of the applicability of the results.

**References**

Alvarez, G. R., & Caregnato, S. E. (2018). Funding acknowledgements in Brazilian scientific output represented in the Web of Science. *Em Questão*, 24, 48-70.

Álvarez-Bornstein, B., Díaz-Faes, A. A., & Bordons, M. (2019). What characterizes funded biomedical research? Evidence from a basic and a clinical domain. *Scientometrics*, 119(2), 805-825.

Alvarez-Bornstein, B., Morillo, F., & Bordons, M. (2017). Funding acknowledgments in the Web of Science: completeness and accuracy of collected data. *Scientometrics, 112*(3), 1793-1812.

Costas, R., & van Leeuwen, T. N. (2012). Approaching the "reward triangle": General analysis of the presence of funding acknowledgments and "peer interactive communication" in scientific publications. *Journal of the American Society for Information Science and Technology*, 63(8), 1647-1661.

Flowerdewa, J. and Li, Y. (2009). English or Chinese? The trade-off between local and international publication among Chinese academics in the humanities and social sciences. *Journal of Second Language Writing*, 18, 1–16.

Franceschini, F., Maisano, D., & Mastrogiacomo, L. (2016). Empirical analysis and classification of database errors in Scopus and Web of Science. *Journal of Informetrics, 10*(4), 933–953.

Gök, A., Rigby, J., & Shapira, P. (2016). The impact of research funding on scientific outputs: Evidence from six smaller European countries. *Journal of the Association for Information Science and Technology*, 67(3), 715-730.

Grassano, N., Rotolo, D., Hutton, J., Lang, F., & Hopkins, M. M. (2017). Funding Data from Publication Acknowledgments: Coverage, Uses, and Limitations. *Journal of the Association for Information Science and Technology, 68*(4), 999-1017.

Hicks, D., Wouters, P., Waltman, L., & Rafols, I. (2015). Bibliometrics: The Leiden Manifesto for research metrics. *Nature*, 520(7548), 429-431.

Huang, M. H., & Huang, M. J. (2018). An analysis of global research funding from subject field and funding agencies perspectives in the G9 countries. *Scientometrics*, 115(2), 833-847.

Huang, Y., Zhu, D., Lv, Q., Porter, A. L., Robinson, D. K., & Wang, X. (2017). Early insights on the Emerging Sources Citation Index (ESCI): an overlay map-based bibliometric study. *Scientometrics*, 111(3), 2041-2057.

Jacob, B. A., & Lefgren, L. (2011). The impact of research grant funding on scientific productivity. *Journal of Public Economics*, 95(9-10), 1168-1177.

Lewison, G and Carding, P. (2003). Evaluating UK research in speech and language therapy. *International Journal of Language & Communication Disorders, 38*(1): 65-84.

Liang, L., Rousseau, R., & Zhong, Z. (2013). Non-English journals and papers in physics and chemistry: bias in citations?. *Scientometrics*, 95(1), 333-350.

Liu, F., Chen, Y. W., Yang, J. B., Xu, D. L., & Liu, W. (2019). Solving multiple-criteria R&D project selection problems with a data-driven evidential reasoning rule. *International Journal of Project Management*, 37(1), 87-97.

Liu, F., Hu, G., Tang, L., & Liu, W. (2018a). The penalty of containing more non-English articles. *Scientometrics*, 114(1), 359-366.

Liu, W. (2017). The changing role of non‐English papers in scholarly communication: Evidence from Web of Science's three journal citation indexes. *Learned Publishing*, 30(2), 115-123.

Liu, W. (2019). The data source of this study is Web of Science Core Collection? Not enough. *Scientometrics*, 121(3), 1815-1824.

Liu, W., Hu, G., & Tang, L. (2018b). Missing author address information in Web of Science—An explorative study. *Journal of Informetrics*, 12(3), 985-997.

Liu, W., Hu, G., Tang, L., & Wang, Y. (2015). China's global growth in social science research: Uncovering evidence from bibliometric analyses of SSCI publications (1978–2013). *Journal of Informetrics, 9*(3), 555-569.

Lundberg, J., Tomson, G., Lundkvist, I., Skar, J., & Brommels, M. (2006). Collaboration uncovered: Exploring the adequacy of measuring university–industry collaboration through co-authorship and funding. *Scientometrics, 69*(3), 575–589.

Moller T, Schmidt M, Hornbostel S. (2016). Assessing the effects of the German Excellence Initiative with bibliometric methods. Scientometrics,109(3), 2217-2239.

Mongeon, P., & Paul-Hus, A. (2016). The journal coverage of Web of Science and Scopus: a comparative analysis. *Scientometrics*, 106(1), 213-228.

Mejia, C., & Kajikawa, Y. (2018). Using acknowledgement data to characterize funding organizations by the types of research sponsored: the case of robotics research. *Scientometrics*, *114*(3), 883-904.

Möller, T., Schmidt, M., & Hornbostel, S. (2016). Assessing the effects of the German Excellence Initiative with bibliometric methods. *Scientometrics*, 109(3), 2217-2239.

Morillo, F. (2016). Public–private interactions reflected through the funding acknowledgements. *Scientometrics*, *108*(3), 1193-1204.

Morillo, F., & Alvarez-Bornstein, B. (2018). How to automatically identify major research sponsors selecting keywords from the WoS Funding Agency field. *Scientometrics, 117*(3), 1755-1770.

Nie, Z. Chinese academic journals should actively participate in internationa academic discourse of power. China Social Science News, 2014-02-19.

Paul-Hus, A., Desrochers, N., & Costas, R. (2016). Characterization, description, and considerations for the use of funding acknowledgement data in Web of Science. *Scientometrics*, 108(1), 167-182.

Paul-Hus, A., Díaz-Faes, A. A., Sainte-Marie, M., Desrochers, N., Costas, R., & Larivière, V. (2017a). Beyond funding: Acknowledgement patterns in biomedical, natural and social sciences. *PloS one*, *12*(10), e0185578.

Paul-Hus, A., Mongeon, P., Sainte-Marie, M., & Larivière, V. (2017b). The sum of it all: Revealing collaboration patterns by combining authorship and acknowledgements. *Journal of Informetrics*, *11*(1), 80-87.

Powell, K. (2019). Searching by grant number: comparison of funding acknowledgments in NIH RePORTER, PubMed, and Web of Science. *Journal of the Medical Library Association, 107*(2), 172-178.

Rigby, J. (2011). Systematic grant and funding body acknowledgement data for publications: new dimensions and new controversies for research policy and evaluation. *Research Evaluation*, *20*(5), 365-375.

Shapira, P., & Wang, J. (2010). Follow the money. *Nature*, *468*(7324), 627-628.

Tang, L. (2013). Does "birds of a feather flock together" matter: Evidence from a longitudinal study on the US-China scientific collaboration. *Journal of Informetrics, 7*(2), 330–344.

Tang, L., Hu, G., & Liu, W. (2017). Funding acknowledgment analysis: Queries and caveats. *Journal of the Association for Information Science and Technology*, 68(3), 790-794.

Ubfal, D., & Maffioli, A. (2011). The impact of funding on research collaboration: Evidence from a developing country. *Research Policy, 40*(9), 1269-1279.

Walsh, J., Lee, Y., Tang, L. (2019). Pathogenic organization in science: Division of labor and retractions. *Research Policy, 48*(1), 444-461.

Wang, J., & Shapira, P. (2011). Funding acknowledgement analysis: an enhanced tool to investigate research sponsorship impacts: the case of nanotechnology. *Scientometrics*, *87*(3), 563-586.

Wang, J., & Shapira, P. (2015). Is there a relationship between research sponsorship and publication impact? An analysis of funding acknowledgments in nanotechnology papers. *PloS one*, 10(2), e0117727.

Wang, X., Liu, D., Ding, K., & Wang, X. (2012). Science funding and research output: a study on 10 countries. *Scientometrics*, *91*(2), 591-599.

Winkelman S, Rots A (2016). Usefulness and dangers of relying on grant acknowledgments in an observatory bibliography. *Proc. SPIE 9910, Observatory Operations: Strategies, Processes, and Systems VI, 99101W*. https://doi.org/10.1117/12.2231668

Wu, X., et al. (2011). The Contemporary Construction of the Chinese System of Academic Discourse. *Social Sciences in China*, 2, 25-27. (in Chinese)

Xie, Q. Journal internationalization and the quality of editors. *Jin Chuan Mei News*, 2013-5.

Yan, E., Wu, C., & Song, M. (2018). The funding factor: a cross-disciplinary examination of the association between research funding and citation impact. *Scientometrics*, *115*(1), 369-384.

Yin, Z., Liang, Z., & Zhi, Q. (2018). Does the concentration of scientific research funding in institutions promote knowledge output? *Journal of Informetrics*, *12*(4), 1146-1159.

Zhao, S. X., Lou, W., Tan, A. M., & Yu, S. (2018). Do funded papers attract more usage?

*Scientometrics*, *115*(1), 153-168.

Zhao, S. X., Yu, S., Tan, A. M., Xu, X., & Yu, H. (2016). Global pattern of science funding in economics. *Scientometrics*, *109*(1), 463-479.

Zhu, J., Hu, G., & Liu, W. (2019a). DOI errors and possible solutions for Web of Science. *Scientometrics*, *118*(2), 709-718.

Zhu, J., Liu, F., & Liu, W. (2019b). The secrets behind Web of Science's DOI search. *Scientometrics*, 119(3),1745-1753.